\documentclass{svproc}
\usepackage{graphicx}
\usepackage{multicol}
\usepackage{footmisc}

\usepackage{cite}
\usepackage{url}

\begin{document}
\mainmatter              
\title{Measurement of non-strange D-meson production and azimuthal anisotropy in Pb--Pb collisions with ALICE at the LHC}
\titlerunning{non-strange D-meson production and azimuthal anisotropy}  
%
\author{Syaefudin Jaelani for the ALICE Collaboration}
\authorrunning{Syaefudin Jaelani for the ALICE Collaboration} 
%
%
\institute{Utrecht University, Princetonplein 1, 3584 CC Utrecht, Netherlands,\\
\email{syaefudin.jaelani@cern.ch}}

\maketitle              

\begin{abstract} 
Heavy quarks are effective probes of the properties of the QGP created in ultra-relativistic heavy-ion collisions. The ALICE Collaboration measured the non-strange D-meson production in Pb--Pb collisions at $\sqrt{s_{\rm NN\,}}=5.02$ TeV.  The in-medium energy loss can be studied via the nuclear modification factor measurement. The measurement of the D-meson elliptic flow, $v_2$, allows us to investigate the participation of the heavy quarks in the collective expansion of the system at low momentum and their possible thermalization in the medium. Furthermore the Event-Shape Engineering technique is used to measure D-meson elliptic flow in order to study the coupling of the charm quarks to the light quarks of the underlying medium.

\end{abstract}

\section{Introduction}
Heavy quarks (i.e. charm and beauty) are effective probes of the Quark-Gluon Plasma (QGP) which is formed in ultra-relativistic heavy-ion collisions. Due to their large masses, heavy quarks are produced in hard scattering processes in the early stages of the collision before the QGP formation, which is about 0.3-1.5 fm/$c$ at the LHC energies \cite{Liu:2012ax}. Thus, they experience the full evolution of the medium and lose part of their energy interactiong with the medium constituents via gluon radiation \cite{Gyulassy:1990ye, Baier:1996sk} or collisional processes \cite{Thoma:1990fm, Braaten:1991jj, Braaten:1991we}. 

The in-medium energy loss can be studied by measuring the nuclear modification factor ($R_{\rm AA\,}$). Information on the transport properties of the medium is obtained by the measurement of the azimuthal anisotropy in the momentum distribution of heavy-flavour hadrons, the elliptic flow $v_2$. The measurement of the D-meson $v_2$ allows us to investigate the participation of low-momentum heavy quarks in the collective expansion of the system and their possible thermalization in the medium.

\section{Non-strange D-meson reconstruction} 
In ALICE, the non-strange D  mesons (D$^{0}$, D$^+$ and D$^{*+}$) are reconstructed in the hadronic decay channels D$^{0}\rightarrow$ K$^{-} \pi^{+}$ (with branching ratio, BR, of (3.93 $\pm$ 0.04)\%), D$^{+}\rightarrow$ K$^{-} \pi^{+} \pi^{+}$ (BR of (9.46 $\pm$ 0.24)\%) and D$^{*+}\rightarrow$ D$^{0} \pi^{+}$ (BR of (67.7 $\pm$ 0.5)\%) \cite{Patrignani:2016xqp}, and their charge conjugates. The decay topologies are resolved via secondary vertex reconstruction thanks to the excellent performance of the Inner Tracking System. Background is reduced by applying topological selections in order to enhance the signal-to-background ratio. In addition, charged pions and kaons are identified using information from the Time Projection Chamber and the Time Of Flight detector. Finally, an invariant mass analysis is used to extract the D-meson yields. The correction for acceptance and efficiency was determined using Monte Carlo simulations based on transport code\cite{Brun:1994aa} which reproduces the detector response. The HIJING \cite{Wang:1991hta} event generator is used to simulate the underlying Pb--Pb events and  D-meson signals were added using PYTHIA6 \cite{Sjostrand:2006za} event generator. The prompt yield of D mesons is obtained by subtracting the inclusive yield from beauty-hadron decays estimated based on FONLL calculations \cite{Cacciari:1998it, Cacciari:2001td}. The V0 scintillators, which cover the pseudorapidity region -3.7 $< \eta <$ -1.7 and 2.8 $< \eta <$ 5.1, provide centrality and event plane angle.
\section{Prompt D-meson nuclear modification factor and elliptic flow}

\begin{figure}
\includegraphics[scale=0.25]{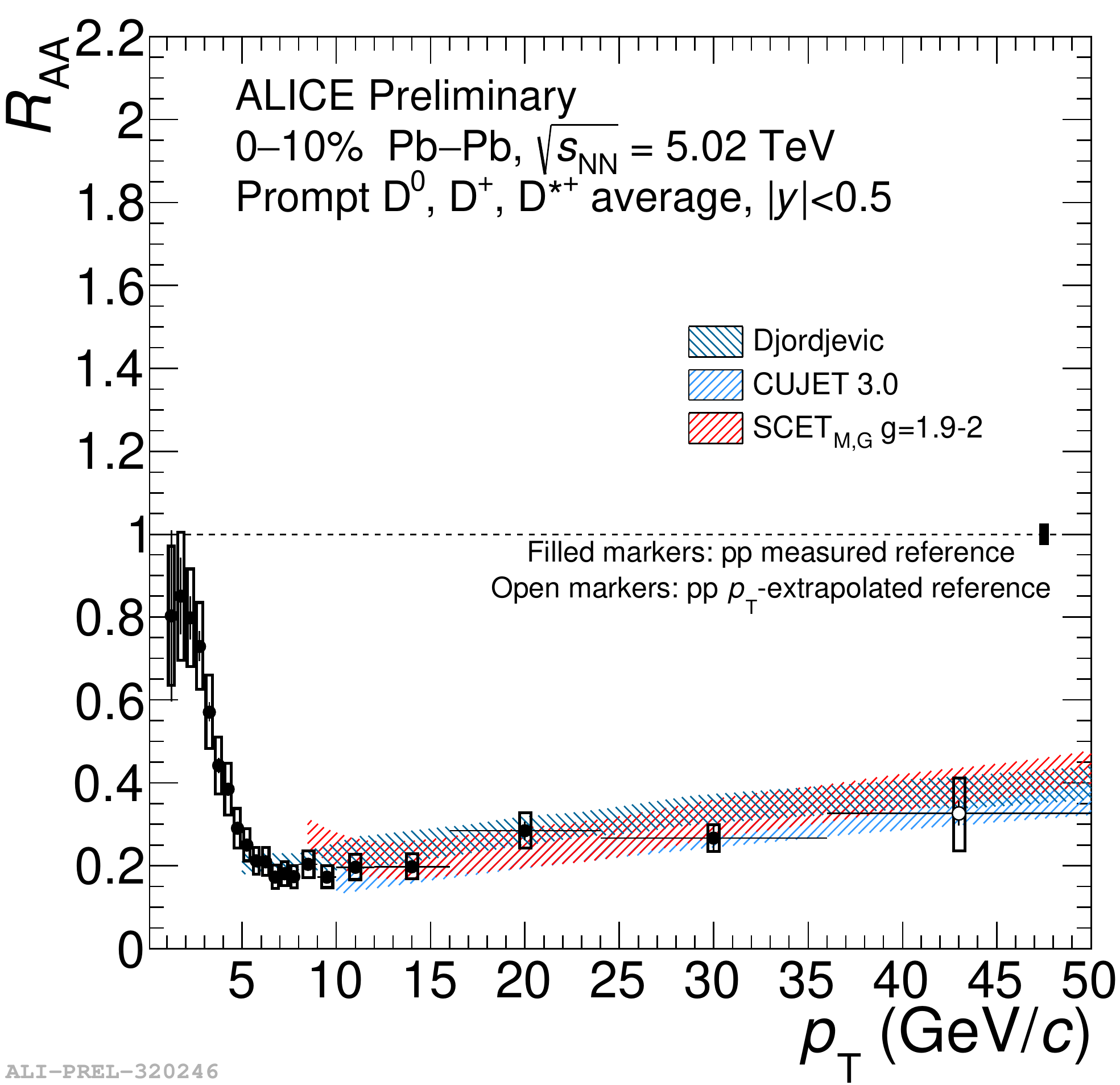}  \hspace{0.5cm}
\includegraphics[scale=0.25]{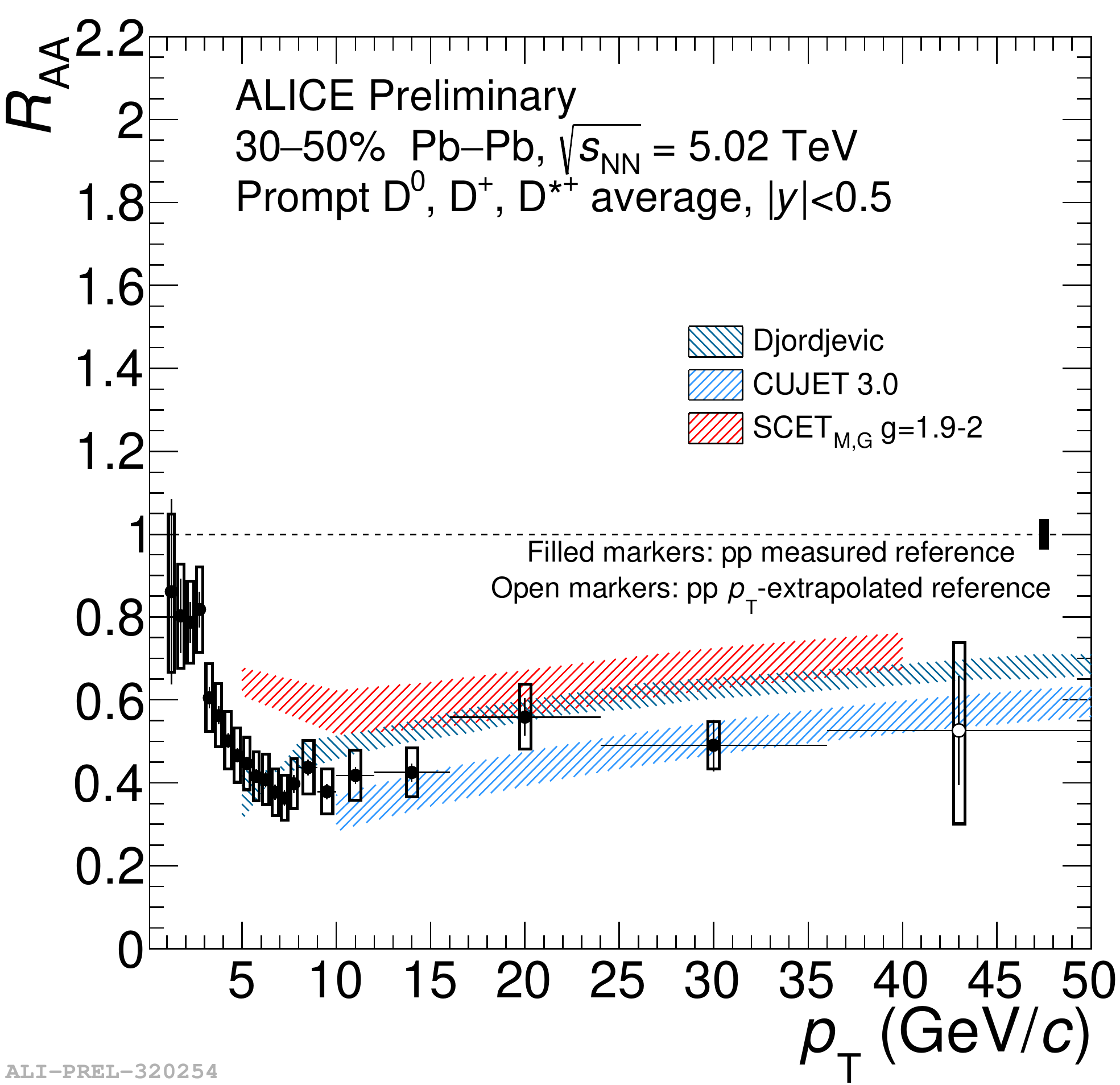} 
\caption{Average non-strange D-meson $R_{\rm AA\,}$ in central 0-10\% (left panel) and semi-central 30-50\% (right panel) events in Pb--Pb collisions at $\sqrt{s_{\rm NN\,}}=5.02$ TeV compared with pQCD model predictions \cite{Xu:2015bbz, Djordjevic:2015hra,  Kang:2016ofv}.}
\label{Fig1}
\end{figure}

The $R_{\rm AA\,}$of non-strange D mesons (D$^{0}$, D$^{+}$ and D*$^{+}$) is measured in two centrality classes, central 0-10\% and semi-central 30-50 \%, in Pb--Pb collisions at $\sqrt{s_{\rm NN\,}}=5.02$ TeV. The $p_{\rm T\,}$-differential cross sections of prompt D mesons in pp collisions at $\sqrt{s}=5.02$ TeV \cite{Acharya:2019mgn} is used as reference. Figure \ref{Fig1} show the average non-strange D-meson $R_{\rm AA\,}$ compared to perturbative QCD model predictions in both centrality classes. The CUJET3.0 \cite{Xu:2015bbz} and Djordjevic \cite{Djordjevic:2015hra} models which include both radiative and collisional energy loss processes, provide a fair description of the $R_{\rm AA\,}$ in both centrality classes for $p_{\rm T\,}>10$ GeV/$c$ where radiative energy loss is expected to be the dominant interaction mechanism.

\begin{figure}
\includegraphics[scale=0.25]{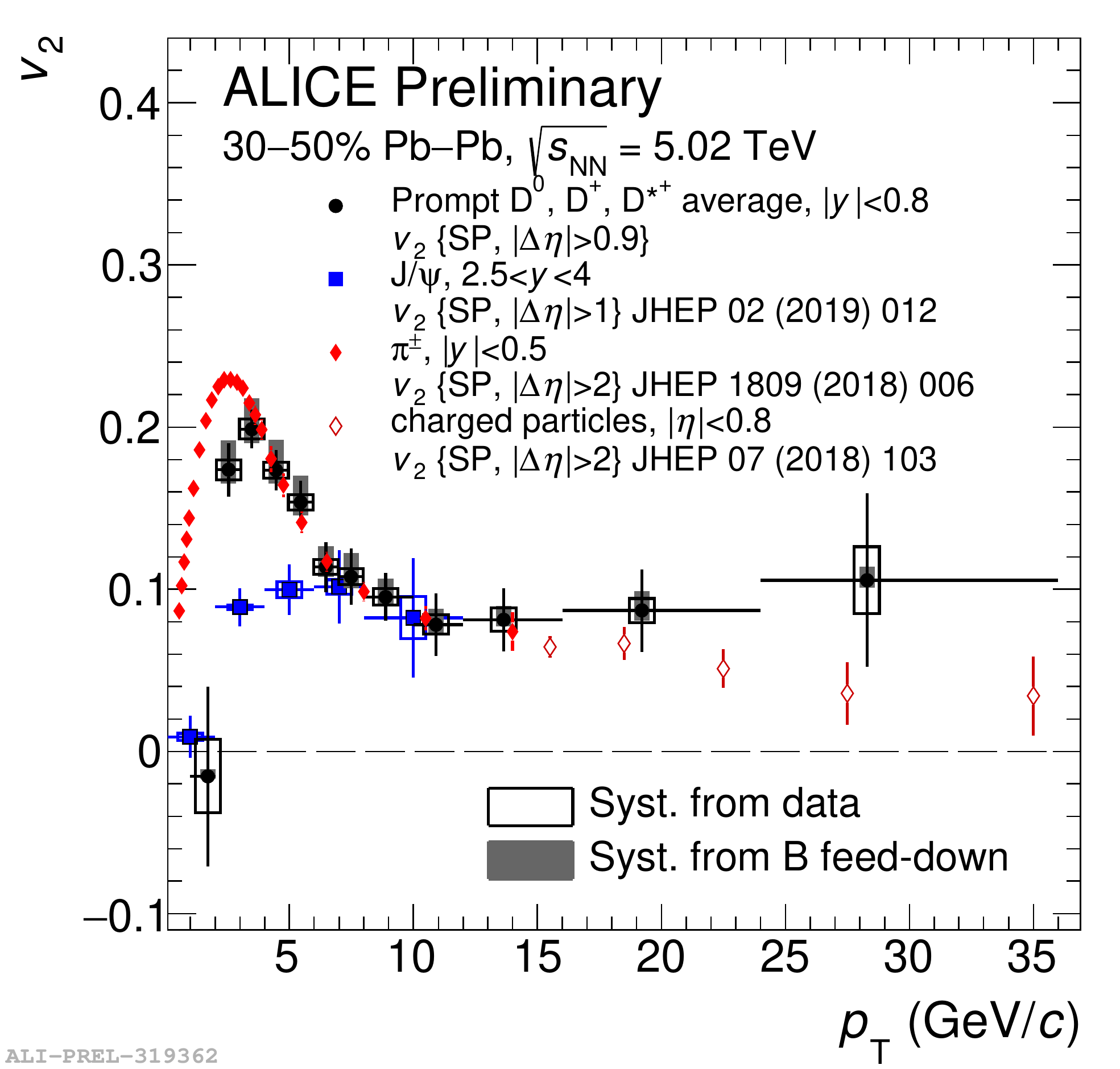} \hspace{0.5cm}
\includegraphics[scale=0.35]{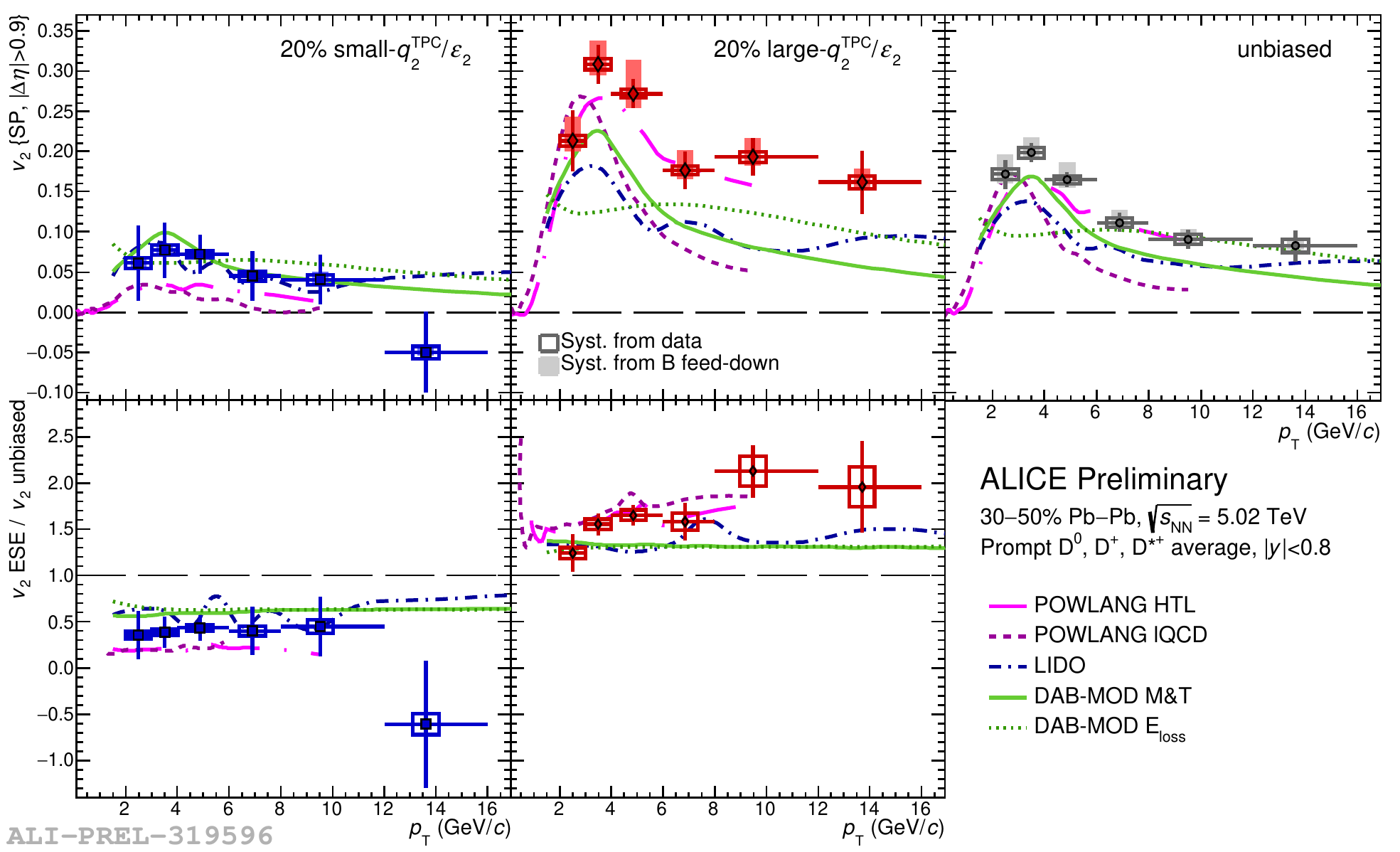}
\caption{Left: average non-strange D-meson $v_2$ in Pb--Pb collisions at $\sqrt{s_{\rm NN\,}}=5.02$ TeV in the 30-50\% centrality class compared to the $\pi^{\pm}$, J/${\rm \Psi\,}$ and charged particles $v_2$. Right: average non-strange D-meson $v_2$ for the small and large $q_2$, and unbiased $v_2$ compared with some of available models.}
\label{Fig2}
\end{figure}

The average non-strange D-meson elliptic flow $v_2$ in Pb--Pb collisions at $\sqrt{s_{\rm NN\,}}=5.02$ TeV in the 30-50\% centrality class is reported in Fig. \ref{Fig2} with the $\pi^{\pm}$, J/${\rm \Psi\,}$ and charged particle $v_2$ in the same energy and centrality class. The $v_2$ of non-stange D-meson are larger than zero for $p_{\rm T\,}> 2$ GeV/$c$ in semi-central Pb--Pb collisions which indicates participation of charm quark in the collective expansion dynamics. 

The Event-Shape Engineering technique was used to inverstigate the D$^{0}$, D$^{+}$ and D*$^{+}$ $v_2$. The second-harmonic reduced flow vector, $q_2 = |Q_{2}| / \sqrt{M}$, can be used to quantify the eccentricity of the events, where $M$ is the multiplicity and $Q_2$ is the second-harmonic flow vector. The events were divided into two groups, small-$q_2$ class (20\% of events with smallest measured $q_2^{\rm TPC}/q_2^{\rm V0A}$) and large-$q_2$ class (20\% of events with largest measured $q_2^{\rm TPC}/q_2^{\rm V0A}$). The average D-meson $v_2$ for the small-$q_2$, large-$q_2$ and for unbiased $v_2$ shows in Fig. \ref{Fig2}, on the right panel, compared with model predictions. The models are based on charm-quark transport in a hydrodynamically expanding medium. The POWLANG HTL\cite{Beraudo:2014boa} model reporoduces well the data for large-$q_2$ and unbiased D-meson $v_2$ below 12 GeV/$c$, while it underestimates the data for small-$q_2$. The LIDO\cite{Ke:2018tsh} and DAB-MOD\cite{Prado:2016szr} models provide better description of the data for small-$q_2$ values, while they underestimate the data for large-$q_2$ and unbiased D-meson $v_2$.


\begin{figure}
\includegraphics[scale=0.25]{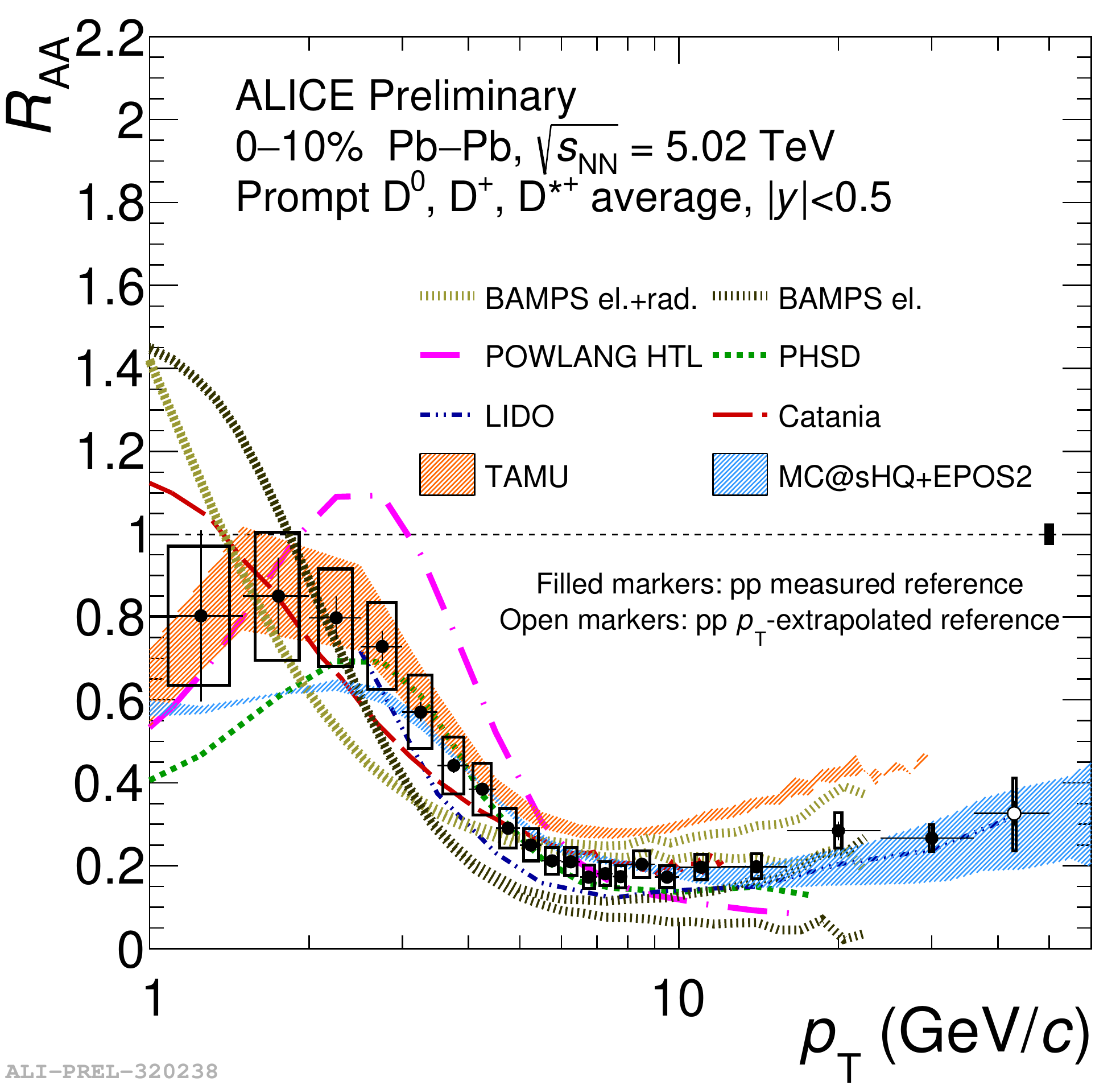} \hspace{0.5cm}
\includegraphics[scale=0.25]{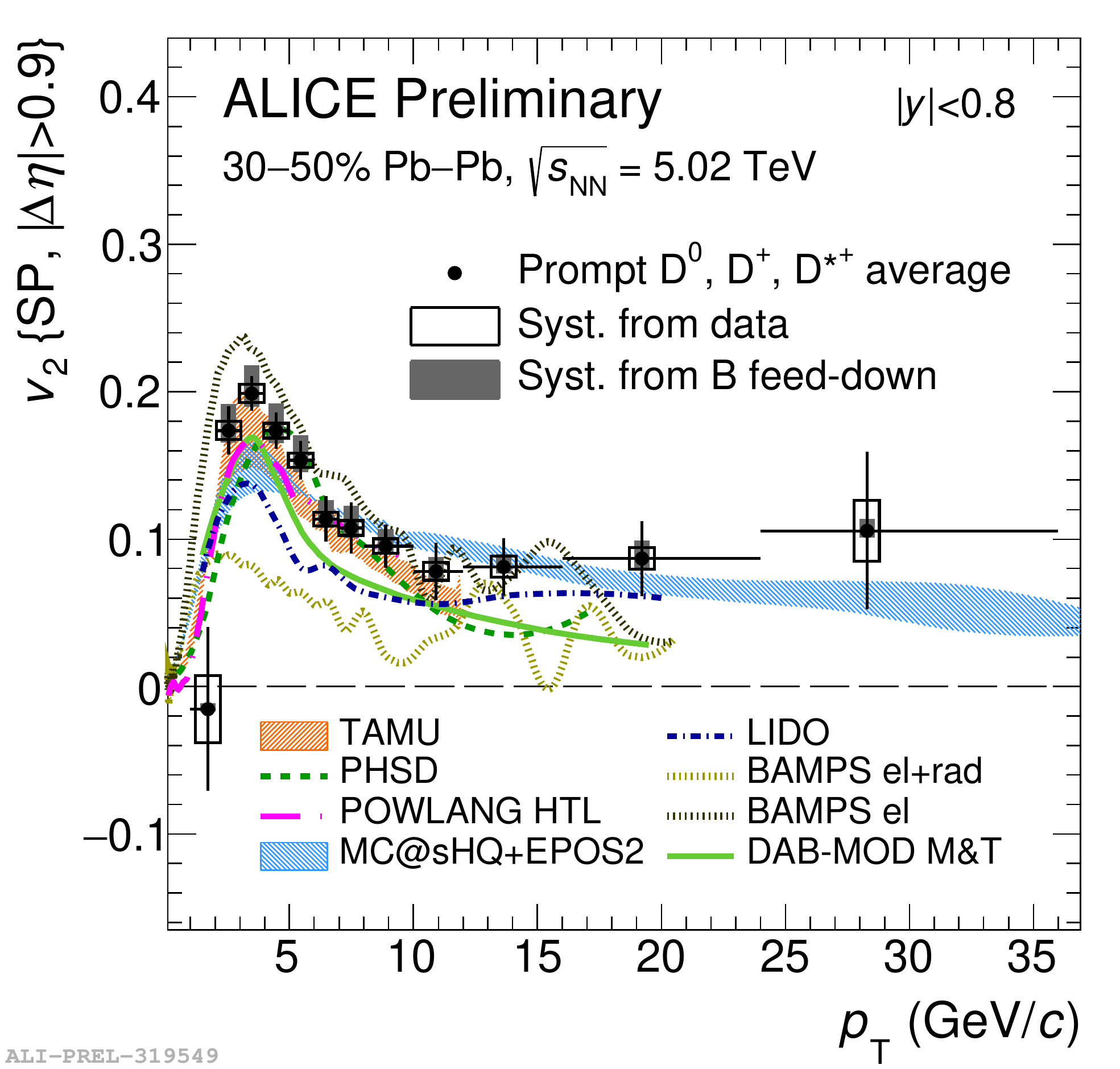}  
\caption{Average non-strange D-meson $R_{\rm AA\,}$ in the 0-10\% centrality class (left) and elliptic flow $v_2$ in the 30-50\% centrality class (right) measured in Pb--Pb collisions at $\sqrt{s_{\rm NN\,}}=5.02$ TeV, compared with the Transport models \cite{Nahrgang:2013xaa,Uphoff:2014hza,Song:2015ykw,Plumari:2017ntm,Ke:2018tsh,Beraudo:2014boa,He:2014cla}.}
\label{Fig3}
\end{figure}

The average $R_{\rm AA\,}$ for non-strange D mesons in the 0-10\% centrality class (left) and $v_2$ in the 30-50\% centrality class (right) are compared with Transport models in Fig. \ref{Fig3}. Most of the models provide a fair description of the data in central events for $p_{\rm T\,} < 10$ GeV/$c$, while POWLANG\cite{Beraudo:2014boa} and BAMPS \cite{Uphoff:2014hza} in which the interactions are only described by collisional (i.e. elastic) processes, show some tension with respect to the $R_{\rm AA\,}$ data points. The TAMU\cite{He:2014cla} model with improved space-momentum correlations between charm quarks and underlying hydro medium, describe well the D-meson $v_2$ for $p_{\rm T\,} < 12$ GeV/$c$. The MC@sHQ+EPOS2\cite{Nahrgang:2013xaa} model provides a fair description of $v_2$, as does PHSD\cite{Song:2015ykw} and TAMU\cite{He:2014cla} for $p_{\rm T\,}<12$ GeV/$c$, while BAMPS \cite{Uphoff:2014hza} model overestimates the maximum flow of $v_2$. In addition, the LIDO\cite{Ke:2018tsh} and DAB-MODE\cite{Prado:2016szr} models describe the shape of $v_2$ but underestimate its magnitude.

\section{Conclusion}
The ALICE Collaboration measured the non-strange D-meson $R_{\rm AA\,}$ and the elliptic flow $v_2$ in Pb--Pb collisions at $\sqrt{s_{\rm NN\,}}=5.02$ TeV. The average $R_{\rm AA\,}$of the non-strange D-meson shows minimum values of 0.15 in the centrality class 0-10\% at $p_{\rm T\,}$ 6.5-8 GeV/$c$ and 0.35 in the centrality class 30-50\% at $p_{\rm T\,}$ 7.5-8 GeV/$c$. The results of the D-meson elliptic flow $v_2$ are larger than zero above 2 GeV/$c$ in mid-central Pb--Pb collisions, which provide information of the collective expansion of the system. In addition, the Event-Shape Engineering technique for the non-strange D-meson elliptic flow $v_2$ was applied.


\end{document}